\documentclass[10pt]{iopart}
\usepackage{psfig}  
\begin{document}\hbadness=10000
\title[Strangeness --- Open Questions]{Quo Vadis Strangeness?}

\author{Johann Rafelski}

\address{Department of Physics, University of Arizona, Tucson, AZ 85721
}
\begin{abstract}
The study of strange and also charmed hadronic particle production  
in nuclear relativistic collisions offers an opportunity to 
explore the physical properties of the deconfined quark-gluon phase.
We survey the recent accom\-plishments and the future directions 
of this research program.
\end{abstract}
\vskip -9cm
Summary talk at the International Conference\\
{\bf Strangeness in Quark Matter},
held in Padova, July 1998\\ 
To appear in: Journal of Physics G
\vskip 8.5cm


%
\section{Unde Venimus} 
\subsection{Preamble}
Nuclear collisions at relativistic energies are the 
experimental tool developed in the past 20 years to form,
explore, and study a locally `color-melted' space-time domain. 
At this time we are assembling  comprehensive experimental evidence 
that a locally color deconfined state indeed is formed 
in present day experiments. I will present  here my personal 
assessment  of the recent developments, current status, 
and a  view of near-future opportunities 
in searching for, and studying the quark-gluon  plasma (QGP) 
using  hadronic flavor observables, strangeness in particular.
My deliberations and conclusions arise from  reports
and overviews presented at the Strangeness 1998 meeting \cite{S98}.  

The experimental and theoretical 
work of interest to us relies on the  idea of 
strangeness enhancement. 
The `heavy' quark flavor, which depending on the energy 
available in the relativistic nuclear 
collision can be strangeness or charm, is a
fingerprint of the structure of dense matter created 
in two distinct ways:
\begin{itemize}
\item[(i)] the overall abundance has to be made nearly entirely 
during the early stages of the collision and thus this yield 
depends on the conditions early on; 
\item[(ii)] the flavor  distribution in the hadronization process
among different final state particles is able to populate
otherwise rarely produced particles, such as strange
antibaryons, which are most impacted 
by the onset of color-deconfinement. 
\end{itemize}

The reach of our current research 
program centered at the SPS accelerator at CERN 
(European Center for Nuclear Research, 
Geneva area across French-Swiss border)
and AGS accelerator at BNL (Brookhaven National Laboratory,  
Long Island, New York)
will be extended within the next two years ten-fold in energy,
with first data emerging from the 
Relativistic Heavy Ion Collider (RHIC), which
is currently being completed at the BNL. By the year 2010 
we will be able to explore entirely 
new  horizons at CERN, where the Large Hadron Collider (LHC) 
will allow to create physical conditions  rivaled only 
by  the Big-Bang. This future time-line leads us through 
at least the period 2015--2020. Looking back to the early 
1980's when first theoretical work has been published 
\cite{Raf80,RH81,BZ82,RM82,Raf84,KMR86}, 
we see that in our work  today 
we have at least as much future as there has been history.

There are yet other considerable research
 opportunities, as in the march towards higher 
energies, where the deconfinement is nearly a sure bet, we 
may have jumped the collision energy 
domain in which the effects related to 
critical behavior could be studied at the conditions that just 
suffice to melt the strongly interacting vacuum structure. At which
energy this situation arises  is presently still a hotly debated topic.

\subsection{The principles}
The physical ideas on which our  research program depends are 
rather simple and hence dependable: if color deconfinement
occurs, the quark-gluon soup will in due course 
equilibrate the flavor abundance between the light 
quarks $u,d$, that are brought into 
the collision  and the initially absent 
heavier and on the laboratory time scale short 
lived $s$ quarks. However, given the hadronic interaction 
scale of $\tau_hc\simeq 10$ fm, the 
lifespan of strangeness,  $\tau_s c\simeq 10^{12}\tau_h$, 
is of course infinite, as is the somewhat shorter lifespan of  charm.
Different mechanisms of flavor pair production have been explored and
it is generally accepted that 
in the deconfined region heavy quarks are primarily produced in gluon 
fusion reactions \cite{RM82,BCDH95,LTR96}. 
For each $s$-quark made, there is its 
$\bar s$ partner, since in strong interactions only quark pairs can
be produced.  The availability of an already prepared 
strange and antistrange quark reservoir at QGP break-up
allows quark clustering into multi-strange final state hadrons without 
`penalty'. In this way, the OZI rule is overcome.
This collective effect, where key components of
a final state particle are made in earlier reactions, is the 
source of our expectation  that we can diagnose the presence 
of QGP using strongly interacting observables. 

One of the keys to the understanding of the nature of 
dense matter provide gluons, the color charged 
cousins of photons. Gluons are the key
distinction between the deconfined quark-gluon phase and the 
confined hadron gas.  They are an important component in determining 
QGP properties, and they are the source of strangeness. Moreover,
gluons play a major role  in the dynamics of the QGP hadronization, 
since  they carry much of the entropy that turns into particle 
abundance. The approximate gluon number  density in QGP is given by: 
\begin{equation}\label{Gab}
\rho_g  \simeq
\left( {T\over {250\,\mbox{MeV}}} \right)^3\mbox{fm}^{-3}\,.
\end{equation}
Because gluons can be created and 
annihilated easily in diverse QCD based strong interactions involving other 
gluons and light quarks, the gluon density is the most capable 
of all QGP components to follow the evolution of the 
exploding/flowing matter closely, maintaining  the chemical equilibrium 
abundance presented in Eq.\,(\ref{Gab}).
The formation time of  QGP, $\tau_{QGP}\simeq 0.2$--0.8 fm,
is widely defined as the time needed for the 
gluon gas to reach initial absolute chemical equilibrium.
Strangeness production by gluon fusion sets in at that time and is
computed  assuming chemical gluon equilibrium.

\begin{figure}[tb]
\vspace*{-8.1cm}
\centerline{\hspace*{6.cm}
\psfig{width=12cm,figure=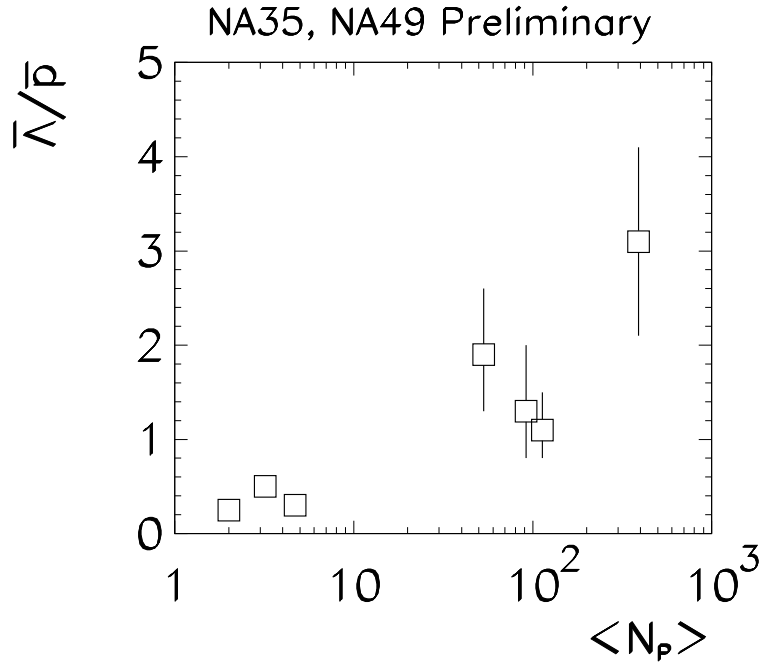}
}
\caption{ 
Preliminary experimental NA35/49
ratio $\overline\Lambda/\bar p$ as function of number of
participating baryons $<N_p>$ for different 
collision systems. From Ref. \cite{NA49Jer}.
\label{pbarNA49}
}
\end{figure}

Hadronic particles seen in the final state can originate from
different production processes; for example, strange hadrons 
may be formed
\begin{itemize} 
\item[---] 
during QGP hadronization, the reaction path of primary interest
here;
\item[---] in initial high energy N--N interactions, 
the always present background in A--A collisions;
\item[---] 
in the  reequilibrating and expanding hadron gas 
state following on the hadronization period, a process that could
obscure the hadronic particle signatures of the QGP state;
\item[---] 
in secondary rescattering from spectator nuclear matter, 
a background we can minimize
by an intelligent choice of triggers and collision partners
\end{itemize}
and should the QGP phase not be formed at all, 
\begin{itemize}
\item[---] during the various (equilibrium and
non-equilibrium) stages of evolution of normal `hadronic gas' (HG)
matter, which is presumed to occur at sufficiently low collision energies.
\end{itemize}
It is thus important to focus our attention 
on the right experimental flavor signature, specifically here
particle family, such as strange (anti)baryons, that is expected to be 
populated predominantly by hadronization of QGP --- in \cite{Raf80}
my concluding phrase says: {\it In the quark-plasma phase we expect a 
significant enhancement of $\overline\Lambda$ production which will be most 
likely visible in the $\overline\Lambda/\bar p$ relative rate.} This 
statement, explained in some detail in Eq.\,(\ref{relsqth}) below, 
 is best compared to the current NA35/NA49 results 
which were recently presented \cite{NA49Jer}, 
see figure~\ref{pbarNA49}. We see that
when the number of participating nucleons $\langle N_p\rangle > 30$,
this ratio is enhanced above the N--N `background'
 where  $\langle N_p\rangle =2$, indeed
there is a 10--30 fold enhancement.  I note that 
the abundance of strange (anti)baryons will not be enhanced significantly 
in confined matter, even if  overall strangeness production  should
increase driven  by some yet unknown mechanisms 
which do not directly  or indirectly invoke the collective 
dynamics present in QGP. 

\begin{figure}[tb]
\vspace*{5.7cm}
\centerline{\hspace*{3.2cm}
\psfig{width=15cm,figure=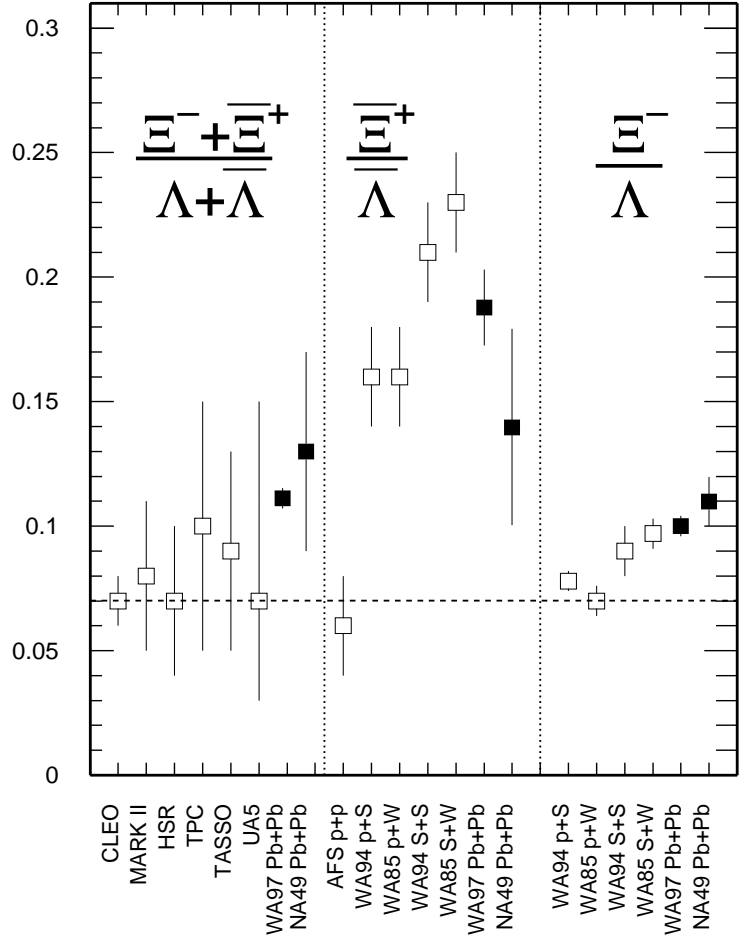}
}
\vspace*{-3.4cm}
\caption{
Ratios of cascades and lambdas and their antiparticles  at fixed  $p_\bot$, 
as `function' of experiment name, ordered such that
central particle multiplicity  $dN/dy|_{\rm CM}$ increases from left to right. 
 Nuclear collision results  are for central 
rapidity   $\Delta y \simeq \pm0.5$, S$p\bar p$S-UA5 high energy 
data ($\sqrt{s}=900,546$\,GeV) are for $\Delta y \simeq \pm2$, 
$pp$-ISR-AFS ($\sqrt{s}=31.5$\,GeV) results 
are most central with $\Delta y \simeq \pm0.2$. Dark squares:  Pb--Pb 158 A GeV 
 results as reported during the
meeting are extrapolated to full $p_\bot$; only 
statistical errors are shown. All other results  
 have slightly varying high $p_\bot$-cuts with the lower
 limit between 1 and 1.4 GeV. 
\label{RSS}
}
\end{figure}

Comparable results are obtained for the ratios of cascades $\Xi(ssq)$ with 
the lambdas $\Lambda(sqq)$  by the NA49 \cite{Gab98} and by the more precise 
experiments WA85, WA94~\cite{Eva98}, and WA97 \cite{Cal98}.
A compilation of these results is shown  in figure~\ref{RSS} 
as a `function' of different experimental projectile and target
combinations. These results were obtained with centrality triggers, assuring
a qualitatively increasing number of participants from left to
right in each of the three sections of figure~\ref{RSS}. 
Dark squares are the recent Pb--Pb results as reported at the
meeting \cite{S98} by the WA97 and NA49 collaborations.  
The central section in figure~\ref{RSS} addresses
the enhancement in the ratio of strange antibaryons.
Combined with the NA35/NA49, $\overline\Lambda/\bar p$ 
results shown above in figure~\ref{pbarNA49},
these results imply a nearly 100 fold enhancement in the $\overline\Xi/\bar p$
ratio over the N--N reaction background, a quite impressive result. Figure \ref{RSS} 
also shows to the very left the production ratios arising in less precise 
measurements involving $e^-e^+$  and $p\bar p$ reactions. We will
discuss the more precise ISR-AFS $p$-$p$ point further below.

We will now show quantitatively 
why this large enhancement in strange antibaryon yield
from QGP phase could be anticipated so long ago.
As discussed, the produced abundances of strange  antibaryons
are not controlled by (small) elementary N--N production cross sections,
but by the quark chemistry occurring in the hadronization process, provided
that the (anti)strange quark abundance is available. While the absolute
particle yields depend on details of the hadronization process, relative
yields of similar anti-baryon abundances are closely related
to the available relative  quark abundance:
\begin{eqnarray}  \label{relsqth}
\frac{\langle \bar s \bar s \bar d \rangle}{\langle 
                         \bar u \bar s \bar d \rangle} 
\simeq 
\frac{\langle \bar s \bar u \bar d \rangle}{\langle 
                         \bar u \bar u \bar d \rangle} 
 &\simeq &  
 \frac{\langle \bar s \rangle}{\langle \bar u  \rangle}
\simeq
\lambda_{\rm u}\frac{\gamma_{\rm s}}{\gamma_{\rm u}}f_{\rm s}\\
&\simeq& 
 1.6 \frac{\overline{\Xi^-}\ldots}{\overline\Lambda+\overline{\Sigma^0}+\ldots}
\simeq 0.56\frac{\overline\Lambda+\overline{\Sigma^0}+\ldots}{\bar p+\ldots} \nonumber
\end{eqnarray}
In the first line in Eq.\,(\ref{relsqth})  we  express the relative quark 
abundance in terms of thermal properties of the system. Here the values of the 
QGP parameters are known today; 20 years ago these have 
been just estimated. For the u-quark fugacity we have
(see figure~\ref{energy} below): $\lambda_{\rm u}\simeq \lambda_{\rm q}
\simeq 1.5$--1.6\,. Here $\lambda_{\rm q}$ is the ($u,\,d$)-light quark fugacity. 
The phase space occupancy ratio $\gamma_{\rm s}/\gamma_{\rm q}\simeq 0.65$ 
as seen in different analysis of the SPS experiments. The relative size of the 
phase space volume of massive to massless quarks is 
(here $K_2$ is a Bessel function): 
$$f_{\rm s}(m_{\rm s}/T_{\rm f})
=0.5 (m_{\rm s}/T_{\rm f})^2 K_2(m_{\rm s}/T_{\rm f})\simeq 0.65\,.$$
In the second line in Eq.\,(\ref{relsqth})
we relate the relative  available quark abundance in the QGP to 
the relative abundance of produced strange antibaryons. This relationship 
applies if these particles are  primarily 
produced in recombination processes, 
including in the available quark number 
the contribution from gluon fragmentation. 
The dots remind us that there are many 
hadronic resonances which can be populated in the hadronization
process. These are taken into account
 with thermally weighted relative strength.  After their fast 
hadronic decay they are contributing to the 
appropriate lowest stable hadronic state. 
We show on the hadronic scale stable 
$\overline\Sigma^0$ state explicitly, since
in all experiments it feeds into the 
$\overline\Lambda$ yield, given its fast
electromagnetic decay. This dilution effect 
is compensated by the factor 1.8 resp. its
inverse, 0.56, which arise from the relative thermal abundance 
of two $\bar u \bar d \bar s$ stable baryons
(the more massive $\overline\Sigma^0(I=1,\, I_3=0)$ is expected to be 
somewhat rarer than $\overline\Lambda (I=0)$). 

We find up to $\cal O$(20\%) precision, inserting in Eq.\,(\ref{relsqth})
the above discussed values of the statistical parameters:
\begin{eqnarray} \label{relsq}
3\,\overline\Xi/\overline\Lambda\simeq
\overline\Lambda/\bar p\simeq 1\,.
\end{eqnarray}
This result is valid provided that there has been, prior to hadronization,
 enough time in the deconfined QGP phase
 to approach absolute strangeness chemical equilibrium. We see that, if 
deconfinement is achieved, multi-strange antibaryons  will be produced in great
 abundance with a specific systematic pattern.  
Moreover, among final state  particles the greatest signal to noise ratio 
has been expected for strange  antibaryons, given  the small background 
arising from N--N reactions. 

Today there are of course much more elaborate theoretical
models, yet this prediction of yesterday has withstood the test of
time, for it relies on very simple
ideas and principles. A quantitative comparison  between 
strange antibaryon production by  (a more elaborate) quark-chemistry 
mechanism  and  a string fragmentation model (tuned to describe
the N--N interactions) 
shows that while quark-chemistry models are agreeing with experiment, 
the string model is not able to reproduce the strange antibaryon 
yields seen in the experiment \cite{Czi98}. 

\subsection{From NN to AA collisions}
Strange antibaryon production has been first explored in the early 1980's 
in  central $pp$--interactions at CERN-ISR (Intersecting Storage Ring)  
by the AFS (Axial Field Spectrometer) collaboration \cite{ISR84},
 where the available
N--N CM-energy was four times greater than it is now under study 
in  nuclear collision experiments
at SPS. The AFS-ISR results at  $\sqrt{s_{\rm NN}}=63$\,GeV were
determined in common fixed interval of transverse momentum 
$1<p_\bot\!\!<2$ GeV: \ 
$\overline\Lambda/\bar p|_{p_\bot}\!\!=0.27 \pm0.02$, \
$\overline\Xi/\overline\Lambda|_{p_\bot}= 0.06\pm0.02$
(see figure~\ref{RSS}), and at slightly higher $p_\bot$:
$\overline\Omega/\overline\Xi|_{p_\bot}\!\!<0.15$ at 90\% confidence. 
A few years later, a similar study was performed at yet 
much higher CM-energies at the CERN-Sp$\bar{\rm p}$S  collider by 
the UA5 collaboration \cite{SppS89} (see figure~\ref{RSS}), with relative 
yields remaining consistently small as given above. 

In view of these emerging 
high energy $pp$-- and $p\bar p$--experimental results, the abundance
anomaly presented in Eq.\,(\ref{relsq}) expected to be
attained at much lower equivalent N--N collision energy seemed
either impossible or, to others, implied that deconfinement would
occur at energies above and beyond studied so far. But
most importantly, there was a hadronic signature worth
to look for experimentally in the nucleus-nucleus
interactions: larger volumes of deconfined hadronic matter, 
when collided even at lower available energies, should 
produce particle abundances with unexpected features,
qualitatively different  from the `elementary' N--N interactions.

The measurement of strange antibaryons 
in nuclear collisions could have in principle be carried out
at the ISR-collider at CERN, where 
studies of inclusive $\alpha$--$\alpha$ reactions were 
already underway \cite{ISRalfa}.   However the ISR, 
capable to deliver heavy ion beams up to 12--15GeV CM-energy per nucleon, 
had been closed to make space for the construction of the Large 
Electron-Positron  collider (LEP). Hence the experimental program 
proceeded at the  SPS, which remained available, being used also 
as an injector for LEP. With  nearly (by a factor $\simeq 1.5$) 
the CM-energy of the ISR, the
SPS could perform better at the time in certain tasks given
the lengthening of the strange baryon weak interaction decay path
by a factor $\gamma\simeq 10$ in view of the center of momentum (CM) frame 
moving with rapidity $y=\cosh \gamma\simeq 3$. However, advances in 
silicon tracking make today a `small' ISR-size collider a very
attractive experimental tool suitable to continue the present day  
fixed target program.

In the past decade strongly interacting flavor probes of dense matter 
have been on the menu of the  CERN-SPS experimental nuclear collision
program. Several experiments, today referred 
to as NA57/WA97/WA94/WA85 and NA49/NA35, explored
  strange hadron production, including strange antibaryons,
  first in 200 A GeV Sulphur (S)  beam interactions
with laboratory stationary targets, including the symmetric S--S reactions, 
and  Sulphur collisions with `heavy' Silver (Ag), 
Gold (Au), Tungsten (W) or Lead (Pb) 
nuclei at 200 A GeV, and, more recently, moving on  
to the Pb--Pb collisions at 158 A GeV \cite{Kin98}. 
These results obtained
at available energy of 8.6--9.2 GeV per participating nucleon
will be soon complemented by strange antibaryon data from 
11 GeV Au-induced experiments carried out at the  BNL-AGS. 

Analysis of the 
global hadronic yields shows a great similarity in the behavior 
of dense matter at the AGS and SPS
energies, despite 4 times higher available CM-energy at the SPS.  
The key  difference between SPS and AGS is that hadronic 
particles produced at the lower AGS energies show 
features characteristic of confined matter chemical equilibration, such
as is a finite non-negligible strange quark chemical potential. In principle, 
this is not in contradiction to possible formation of deconfined 
baryon rich quark matter at AGS energies, since hadronic abundance 
equilibration could  be just a final state effect. Moreover, since
for AGS energies it is expected that the specific entropy per baryon 
in confined and deconfined matter is very similar, there
is   to this date no clear evidence for, or against, QGP
formation in the dense baryon rich matter fireballs made at AGS. 
The forthcoming measurement of strange antibaryon production should 
resolve this issue in the near future. 

The measurement of strange antibaryons is  not  easy,  and 
experiments need to be designed for the
task as  the relatively 
rarely produced  strange antibaryons are
literally buried in much more abundant mesons. Ways have to be 
designed to eliminate that background without loosing the acceptance,
which is required to observe mesons emerging from the
 self-analyzing decays such as 
$\overline\Xi\to \overline\Lambda+\pi\to N+\pi+\pi$. 
An illustration of the experimental problems we experience is 
available when we look at the precision of the measurement of the 
ratio $\overline\Lambda/\bar p$. The preliminary
NA49 result for Pb--Pb collisions \cite{NA49Jer}, 
see figure~\ref{pbarNA49}, has a 
precision of 30\%, so within 3\,s.d. one could claim that the ratio
is negligible, even though quite evidently in Pb--Pb collisions
it is significantly  greater than unity, defying the
normal hadron production  rules. One of the problems of 
measuring nucleons and anti-nucleons is that they do not self-analyze, 
which  the weakly decaying  strange baryons and antibaryons do.
Moreover,  the negatively charged antiprotons are at CERN energies produced 
less abundantly  than   K$^-$. Given the relativistic boost in fixed
target experiments the proper identification of these particles poses a
practical challenge. 

\subsection{Strangeness enhancement}
The validity of the strange particle 
signatures depends on the ability to produce high strangeness abundance
in the dense matter. Considerable effort has therefore been devoted to the
understanding of the kinetic mechanisms of flavor production.
Kinetic theory studies of flavor production in QGP  have 
shown that both the total strangeness yield,  and thus the related
emergence of an enhanced strange antibaryon yield \cite{KMR86,LRT96b}, 
are to be expected, and that the dominant strangeness production mechanism in
hadronic matter is based on gluon induced reactions \cite{RM82,BCDH95,LTR96}. 

Whichever of the microscopic mechanisms one adopts for computation of the
strange flavor production in the yet unknown form of high density nuclear
matter that has been generated in relativistic nuclear collisions, 
one can identify the different factors controlling the yield in a 
rather model independent way. Consider two as yet
unidentified constituent parts of centrally interacting nuclei, $A$ and
$B$, producing strangeness in individual collisions. 
 The rate of production per unit of time and
volume is given by
\begin{eqnarray}
\left(\frac{dN_s}{dVdt}\right)=\,
   \langle\sigma^s_{AB}v_{AB}\rangle\rho_A\rho_B\ .
\end{eqnarray}
Since  $\rho=N/V$, the specific strangeness yield per final state hadron is:
\begin{eqnarray}\label{snumber}
\frac{N_s}{n_{\rm h}}\simeq \frac{N_A}{n_{\rm h}}\cdot 
 t\, \cdot \langle \frac{N_B}{V} \cdot\sigma^s_{AB} v_{AB}\rangle\ . \label{NS}
\end{eqnarray}
The relative particle abundance factor $N_A/n_{\rm h}$ is nearly  independent of 
the internal structure in the dense matter fireball: the number of
flavor producing particles  $A$, which could be gluons and
quarks, or could be pions, is a  measure of the 
final hadron multiplicity $n_{\rm h}$, except if new mechanisms 
are yet discovered for entropy 
production during the final state evolution. Similarly, 
the lifespan of the dense phase cannot depend decisively on the internal
structure of the fireball, 
and thus as seen in Eq.\,(\ref{snumber}), enhancement of strangeness 
production will be due primarily to the following two factors:
\begin{enumerate}
\item smaller effective volume $V$ per particle (higher density), and/or
\item enhanced microscopic cross section (e.g. 
dissolution  of production thresholds).
\end{enumerate}

Both these effects occur naturally in the deconfined phase. 
The  second one has  been  postulated 
in order to generate strangeness enhancement in normal, confined hadronic matter. 
Within such an ad-hoc scheme  the challenge for the inventor is then 
to demonstrate some other physical effect that would arise 
from such new and yet undiscovered confined hadronic matter phenomena. 
Perhaps equally relevant to the assessment of such proposals
 is the fact that one has to look for more
than `strangeness enhancement' with flavor observables, 
and in particular the (relative) yields of strange antibaryons originating in 
novel forms of confined matter should be also compared with the experimental 
data.

\section{Ubi Sumus} 
\subsection{Potential range of current experimental program}
\begin{figure}[tb]
\vspace*{-0.7cm}
\centerline{\hspace*{1.7cm}
\psfig{width=13cm,figure=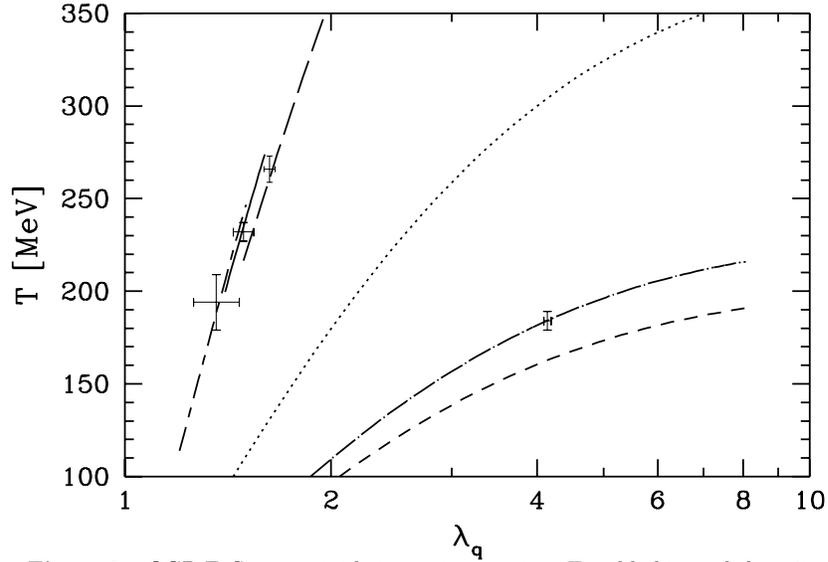}
}
\vspace*{-0.8cm}
\caption{ 
QGP-EoS constraint between temperature 
$T$ and light quark fugacity $\lambda_{\rm q}$ for a given fireball
energy content per baryon $E/B$ appropriate for the AGS and SPS
collision systems. Right to left: 2.3 (Au--Au), 2.6 (Si--Au),
4.3 (A--A), 8.6 (Pb--Pb), 8.8 (S--PB/W) and 
9.6 (S--S) GeV. 
The Fermi($u,d,s$)/Bose($G$) liquid EoS is 
used here to evaluate the QGP properties, with
thermal particle masses $m_i^2(T)=(m_i^0)^2+2T^2$\,, and 
perturbative QCD interactions for gluons and $u,d$ quarks
with $\alpha_s=0.6$\,. For theoretical details
see \protect\cite{LRT96b}.  \label{energy}
}
\end{figure}

An illustration of the potential QGP 
coverage offered by the CERN-SPS and BNL-AGS accelerators
is presented in figure~\ref{energy}, where  i show the 
kinematic constraint between the statistical
parameters temperature $T$ and $\lambda_q=e^{\mu_q/T}$,
the chemical light quark fugacity in an 
equilibrated QGP  phase. To obtain these results \cite{LRT96b}, 
perturbative QCD equations of state  (EoS) were 
evaluated in first order using $\alpha_s=0.6$, and 
furthermore non-perturbative effects
are taken into consideration by allowing for thermal quark and gluon masses:
$m^2=m_0^2+k(\alpha_s)T^2 $, where $k=2$=Const. was chosen. 
At this time it is not 
known how well such treatment of the QGP describes the transition
to the confined HG phase. On the  other hand, the two inter-dependent  parameters 
$c,\alpha_s$ as chosen yield properties of QGP at $T\simeq 250$ MeV consistent
with the observed nuclear collision results. Thus in that domain, assuming 
deconfinement, one can consider QGP EoS so determined 
to be adequate.

 For each  line shown in figure~\ref{energy}
 the QGP energy per baryon $E/b$ is set equal to the available
kinematic energy content per baryon in the collision: 
\begin{enumerate}
\item the dashed line bottom right is for the center of momentum (CM) per baryon 
energy $E/b=2.3$\,GeV, corresponding to 11.2\,GeV A projectile on fixed target 
Au--Au interactions at AGS;
\item the dot-long-dashed line next to it is for $E/b= 2.6$\,GeV,
corresponding to 14.6\,GeV A Si--Au interactions at AGS;
\item the dotted line in the middle is for $E/b=4.3$\,GeV, corresponding to
the low energy 40 A GeV limit of the SPS A--A  runs scheduled for autumn 1999; 
\item the long dashed line is for the 158 A GeV Pb--Pb collisions at SPS,
 yielding $E/b= 8.6$\,GeV\,;
\item the solid line just next to it is for the 200 GeV asymmetric S--Au/Pb/W  
collisions, which give a slightly higher CM-energy content per nucleon,
 $E/b=8.8$\,GeV, and 
\item the short-long dashed line is for the symmetric 200 GeV S--S collisions which
in the CM frame yield is $E/b= 9.6$\,GeV.
\end{enumerate}
Along these lines many properties of the system vary, such as entropy per 
baryon, pressure, baryon density. This variation corresponds to the variation in
impact parameter and size of the colliding system: when the colliding nuclei 
are heaviest, there is less transparency, the baryon number is more compressed,
the pressure of collision is higher, one reaches more extreme conditions 
moving to higher $T,\,\lambda_q$. The same trend appears with 
increasing centrality of the collision, i.e., decreasing impact parameter. 

The crosses in figure~\ref{energy} represent  the  conditions 
reached in the respective collisions systems,  based on chemical 
freeze-out analysis for $\lambda_q$, and the inverse slope of
transverse energy particle spectra for $T$\,. The rational for 
this estimate  of the initial conditions in the dense 
QGP phase  formed in the nuclear collision \cite{LRT96b} is as follows:
\begin{itemize}
\item[$T$:\ ] The $m_\bot$ inverse slope, rather than e.g.\,$T_{\rm f}$,
the freeze-out temperature, is employed, since one finds in
schematic models that a qualitative relationship remains between
the initial temperature and the high, $m_\bot>1.5$ GeV, inverse slopes. 
Intuitively this can be easily understood: when the transversely oriented 
flow of matter sets in at high compression, it acquires thermal
energy which is converted to collective kinetic energy. 
Along with energy conservation this roughly assures the relationship we need.  
However, both the nature of the transverse
flow, and the way  the inverse slope is determined introduces considerable
model dependence. I thus believe that the inverse slope may be easily 10-15\% 
different from the conditions at the onset of flow. On the other hand, the
relative change in the inverse slope is much less model dependent as the 
collisions system is varied or the reaction energy changed. Thus the systematic 
change of the magnitude of the inverse slope with the size of the collision system 
shown in figure \ref{energy} is telling us that significantly higher iniital 
temperatures are reached in Pb--Pb reactions compared to S--S reactions at 
the current SPS energies. 
\item[$\lambda_{\rm q}$:\ ] For isentropic (constant entropy, or equivalently
constant entropy per baryon) expansion of the fireball to freeze-out 
condition the chemical fugacity undergoes negligible variation. This 
can be easily understood, since the specific entropy per baryon of massless
quarks (massless on the scale of temperature) does not depend on a
 dimensioned parameter, thus it is just a 
function of $\lambda_{\rm q}$, {\it i.e.} $S/b=f(\lambda_{\rm q})=$Const., and
hence $\lambda_{\rm q}=$Const. for the isentropic expansion. A small 
correction at the level of a few percent
is found due to the change of the quark-quark interaction strength, 
which varies logarithmically. Also, the strange quark mass is not 
negligible near freeze-out, but this strangeness effect is largely 
compensated by the entropy production
associated with the formation of $s\bar s$-pairs.  
\end{itemize}
As shown in figure~\ref{energy}, the SPS
points cluster to the left and rise with the size of colliding system, 
from bottom to top: S--S, S--W/Pb, Pb--Pb.
The AGS result shown to  the right in the figure for  Si--Au case,
falls in principle into a different category, since the strange chemical 
potential measured in the reaction does not vanish, yet it falls onto
the QGP constraint line. 

Inspecting the SPS results we see that a variation in the atomic number $A$ 
of the colliding nuclei, which can be nearly equivalent to a choice of different
 centrality of the reaction, will  introduce another data point along these three 
lines shown. In order to fill the big gap 
in-between SPS and AGS, a 40 A GeV  run is scheduled for 1999 at the SPS. Thus
within the forthcoming 18 months we will be able to bridge (see 
dotted line in figure~\ref{energy}) the energy
gap between CERN and AGS data. Such a more complete coverage of the
statistical parameter range  is the primary physics motivation
to reduce SPS beam energy.

\subsection{Experimental progress}
Do any of the recent strangeness results suggest
the presence of new physics, or can we deal with these data on the basis
of known phenomena, without invoking the local color deconfinement?  
A clear answer to this question is,
in my opinion not available yet, but a number of highly noteworthy 
experimental and theoretical results are
pointing in the direction of QGP, we have here discussed in depth the
enhancement of antibaryons which has been observed at CERN-SPS 
in the way it has been predicted. Thus, if there
was other convincing supporting evidence, e.g. from dilepton or direct photon
observables, this conclusion could probably be reached today. Moreover, the
widely discussed topic of charmonium suppression is not helping here, since 
there seems to be a great difference in the physics result comparing S and Pb
induced reactions, which we do not really notice in the strange particle 
abundances.

In addition,  the experimental strangeness results have not yet
been analyzed in full. The sheer abundance of experimental hadronic 
particle data  is very large,  and it takes considerable effort  
to obtain precise hadronic abundances and spectra. 
For example, in the past 18 months, the 
WA97 collaboration has analyzed only a fraction 
of their data tapes obtained in  the autumn 1996 run involving 
Pb--Pb 158 A GeV collisions at CERN-SPS~\cite{Cal98}: their analysis 
is reaching 40\% for multi-strange (anti) baryons 
$\Xi,\,\overline\Xi,\,\Omega,\,\overline\Omega$, 
but barely scratched the analysis of $\Lambda$ and $\overline\Lambda$ 
yields. The NA49 collaboration continues to review their
initial 1995 and 1996 $\overline\Lambda$  and $\Lambda$ results
and has presently announced  progress in understanding the 
impact of the cascading decays from $\Xi$ and $\overline\Xi$\,, 
which populate the central yields of $\Lambda$ and $\overline\Lambda$ 
particularly strongly, given the large size of the NA49-TPC set-up \cite{Mar98}.

\begin{figure}[tb]
\vspace*{-9.cm}
\centerline{\hspace*{2.9cm}
\psfig{width=12cm,figure=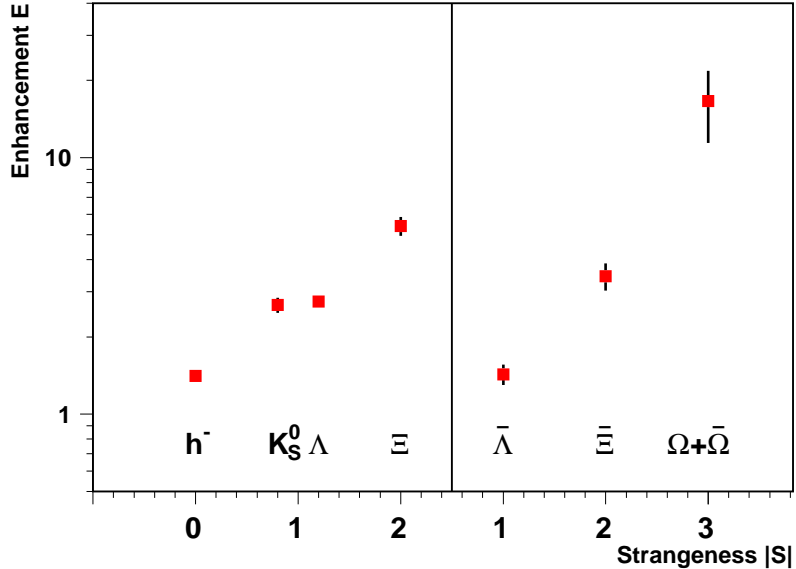}
}
\vspace*{-0.5cm}
\caption{ 
Strangeness enhancement versus strangeness, WA97 collaboration 
\protect{\cite{WA97LietPad}.} \label{Liet11}
}
\end{figure}
%

Returning to the possibility of QGP formation, we note
the strangeness enhancement effect has been studied now in considerable 
detail, and there is a well reported evidence assembled 
over the past 10 years in all the pertinent BNL and CERN experiments.
Without presenting it here in every detail, I note
that  in several slightly varying
definitions of the word, this enhancement is reported to be 
a factor  $2.5\pm 0.5$\,, as it has been  predicted if QGP is formed. 
Moreover, this enhancement is consistently rising with the 
multi-strangeness of hadrons. This is most easily seen in the recent 
results presented by R. Lietawa for the WA97 collaboration \cite{WA97LietPad}, 
see figure~\ref{Liet11}, where the different enhancement factors
obtained in the same experiment (same trigger centrality, same data 
analysis method) are shown.  We see in figure~\ref{Liet11}
the enhancement over the  expected 
yield, with the basis provided by the  N--A reactions
measured in the same experimental setup. The extrapolation from
A--A reactions to N--A reactions
is made implementing a linear scaling of 
particle yield with the number of participant
nucleons, a point carefully documented by the WA97 collaboration in terms of 
the trigger dependence of particle yields. This 
part of the analysis shows that the A-scaling of 
strange particle production is assured, when the 
number of participants exceeds about 100.
Thus, somewhere on the way from a few participants 
(N--A interactions) to triggered
central A--A interactions, the yield has to rise in a  
nonlinear fashion with the participant number
and the strangeness enhancement here presented arises.  
Note that in figure~\ref{Liet11}  the 
particle abundance  enhancement increases significantly as we consider the 
production of particles, which require assembly of constituents not brought
into the reaction by colliding nuclei. 

Figure~\ref{Liet11} also shows
enhancement of negative hadrons (factor 1.5) and single-strange  hadrons 
(factor $\simeq 2.8$  for K$_s,\,\Lambda$ and factor $\simeq 1.5$ for 
$\overline\Lambda$). Thus the  entropy and strangeness yield
are both increasing in parallel in Pb--Pb  reactions, again as would be expected 
if melting of the color bonds of quarks occurs for both light and strange quarks. 
But the crucial point seen in figure~\ref{Liet11} is that
the yield increase of multi-strange hadrons is in general much stronger,
 reaching a factor of 4.5 for $\Xi$ and above 10 for the triple strange Omegas. 
This effect is the anticipated and specific
signal of formation and breakup of a deconfined 
space-time region, and has so far eluded models that rely on confined matter.
In conclusion, we see a systematic pattern of strange particle enhancement 
in figure~\ref{Liet11} and note that at present the only model capable to 
reproduce this result is QGP fragmentation-recombination hadronization.

The current status  regarding relative abundances at central 
rapidity of cascades and lambdas and their antiparticles 
is  displayed in figure~\ref{RSS}\,, while the relative yields of anti-lambdas
and antiprotons are shown in figure~\ref{pbarNA49}.
In figure~\ref{RSS}, the dark squares are 
the recent Pb--Pb results as reported at the
meeting \cite{Gab98,Cal98}.  These new results refer to the full 
range of $p_\bot\in (0,\infty)$, while the earlier
data (e.g., of collaborations WA85 and WA94) refer to ratios for 
$p_\bot > 1.2$ GeV. This explains in part the reduction in value of 
some of the ratios shown. The other effect contributing to the 
reduction of $\overline\Xi/\overline\Lambda$-ratio is the Coulomb 
effect in Pb--Pb collisions, which has noticeable
impact on relative strange particle yields  from  deconfined matter. 
Note also the $pp$ and $ee$ interaction results in  figure~\ref{RSS}\,, 
which offer  comparison with the backgrounds 
from elementary interactions. Current WA97 data offer for  the first time 
solid proof (by many standard deviations)
of the enhancement by a factor of 3  for 
$\overline\Xi/\overline\Lambda$, since both
the results from ISR-AFS-pp interactions 
and the WA97 results from Pb--Pb interactions
are sufficiently precise. This enhancement 
is even more remarkable, considering that 
the available energy in the Pb--Pb collisions 
has been four times smaller than the 
energy available at the ISR N--N interactions.  

The  AGS experiments were initially focused on other `strange' issues, and
in particular many resources have been vested to seek strange
nuclear matter. Only recently the study of strange antibaryons as 
a possible signature of the deconfined state, has commenced, for there are 
strong indications, e.g. in the E802/E859/E866 experiment series, that 
the yield of $\overline\Lambda$ exceeds the yield of $\bar p$ at AGS 
energies. We have seen at the meeting, that there is a continued 
and considerable effort devoted 
in experiments E917 \cite{Gan98} and E895 \cite{Chu98} to measure  
systematically  strange particle yields as a function of energy, 
up to 11~A GeV. It is an interesting race against time, for the
completion and availability of RHIC is likely  to overpower AGS
experimental efforts in a not too distant future.

\subsection{Hadrons from QGP}
If quark-gluon plasma is produced in nuclear collisions, a major
theoretical challenge is the understanding of the hadronization 
process. In principle, 
production and emission of particles can occur throughout the 
evolution of the dense matter fireball. However, it has been
generally assumed that the bulk of particles is produced in the
final moments of evolution when the temperature of dense matter
sinks below the deconfinement condition.  
The cooling of matter is not necessarily 
a result of energy radiation, generally it is believed to be 
primarily  due to the transfer of local thermal 
energy to the collective flow of matter. Such a collective expansion 
process can be nearly  entropy conserving,
since  for $S\propto VT^3$ [or better $S\propto\int d^3xT^3(x)$] 
the increase in system size is compensated by a
corresponding decrease in $T$.

To proceed, we need to clarify the understanding of the different 
 local chemical equilibria introduced earlier \cite{KMR86}. We  
distinguish two cases for their different reaction time scales. The 
slower {\it absolute} chemical equilibration requires that the
particles produced build-up to fill the available phase space; for the 
faster, {\it relative} chemical equilibrium, a  weaker requirement suffices,
namely the redistribution of some property (here quark flavor)
among different carriers (particles) according to the relative phase
space size.  A sample of alternative evolution scenarios that 
could be responsible for the production of the final state hadrons includes:
\begin{enumerate}
\item If the hadronization 
temperature is low (say below 145\,MeV), it can be safely assumed
that there will be little, if any, subsequent change 
in the hadron abundances. Chemical equilibrium, in which quark and gluon
abundances are fixed to the Stefan-Boltzmann limit should not be presumed,
neither for the source of the hadrons, nor for the hadron yields after hadronization.
In this case of `low'  hadron formation  temperature,  hadronization 
is also the chemical freeze-out, and the study of the chemical freeze-out 
conditions can reveal valuable information about the QGP phase. 
This seems to be the situation we encounter at SPS energies. 
\item If hadron production from the deconfined phase were to occur at 
conditions that are more  dense (in terms of baryon density, 
or particle density in general, synonymous with higher temperature), 
chemical re-equilibration among confined final state hadrons should
occur, erasing the eventual particle abundance signature of 
the primordial QGP phase; this is possibly the situation at AGS energies.
\item The possibility of explosive and continuous disintegration cannot
be ruled out at present. In this scenario, beginning at high 
temperatures QGP phase decays successively  
by emission of free-flowing particles.
In that conditions we can use hadronic particles again to establish the
properties of the plasma phase. For isentropic fireball evolution, which 
in the QGP phase means that the quark fugacities are 
unchanged, a consistent determination of chemical parameters is also 
possible in this case, using final state hadron abundances. 
\end{enumerate}
Since QGP chemical properties are rather characteristic, e.g., the 
strange quark chemical potential is nearly zero in the deconfined phase,
analysis of hadron abundances can tell us if a re-equilibration process
has been occurring. The experimental data at SPS energies 
favors, as I see it, a scenario in which 
no chemical re-equilibration did occur after hadronization. This means that
hadronization/chemical freeze-out are the same process.

Even if re-equilibration were to
occur, dynamical calculations show \cite{KMR86}
that it is highly unlikely that the number of 
strange quark pairs  changes significantly once, speaking in relative terms, the 
low density post-freeze-out phase has been established. One can thus
infer strangeness abundance and phase space occupancy conditions present
at QGP hadronization rather precisely from the final state hadron abundances. 
To derive QGP strangeness occupancy we need to consider that the  phase space density of strangeness  in QGP and HG phases is different. Therefore,  
the phase space populations must be appropriately adjusted to infer from
the observed final state yield of strange quarks per baryon number 
the conditions in the earlier QGP phase. Moreover, several
chemical analysis of the data, which reported that the strangeness
phase space was saturated up to about 70\% in the final hadronic phase, have 
indeed been reporting the {\it ratio of strange to non-strange} phase-space 
saturation. The non-strange quark occupancy is at 150--200\% of the equilibrium,
as is noted by the excess of mesons (excess entropy). Allowing for this effect 
we see that the strangeness phase space is  overpopulated in the HG phase, but 
not in QGP. Without invoking QGP formation it appears as if 
the chemical strangeness equilibrium were approached from above, a situation 
difficult to justify.  The parallel overpopulation of 
the light  quark phase space is also in contradiction with the HG phase
hypothesis, but is also expected to arise if QGP is formed: 
the glue degree of freedom must fragment into quarks in the hadronization 
process in order to preserve the enhanced entropy originating in the
melted color bonds. 

As particle emerge from the QGP hadronization, not only their abundance
but also their spectra will be unusual. Indeed, 
some bold conclusions about the origin of the strange antibaryons
can be inferred from the remarkable experimental fact that the 
transverse mass spectra are so similar for particle-antiparticle pairs 
(e.g.,  $\overline\Lambda$--$\Lambda$, $\Xi$--$\overline\Xi$), 
as seen  in results of the precise measurements made by 
experiments NA49 and WA85/WA94/WA97: 
\begin{enumerate}
\item The thermal equilibrium in heavy ion collisions 
is relatively well established; for very different spectra 
should be arising from hadron based reactions
(for strange baryons associate production  N+N$\to\Lambda+$K,
 direct pair-like production  for antibaryons) as is seen in
many microscopic models simulating the nuclear collision process. 
\item The exponential shape of the $m_\bot$-spectra with a common 
inverse slope  implies further that
these particle pairs, or the building blocks from which they are made
in a coalescence picture, had reached well thermalized condition. 
\item The shape identity  of the transverse mass
 spectra of these pairs implies that they have either been dragged in the 
same manner by the flowing hadronic matter, or that they were emitted
by a flowing surface source and reached the detector without much 
further interaction.
\item Since the drag forces of the flowing matter can be expected to be 
greatly different for the $\overline\Lambda$--$\Lambda$ particle pair, and 
since it can be today  subsumed that there is considerable 
transverse flow with $|\vec v_\bot|\propto 0.5\,c$ at the time of particle
freeze-out, one is driven to 
the conclusion that these strange baryons and antibaryons were not 
`dragged' along in confined  matter, and thus must have been 
formed in coalescence  of flowing, thermally equilibrated deconfined matter. 
\end{enumerate}
We see that not only the yields of strange
(anti)baryons, but also  the details of their spectra are relevant
and will contribute significant information about the production mechanism 
and the possibly deconfined nature of the hadron source.

\subsection{Theoretical directions}
There are  many theorists working on flavor observables
and the theoretical tools are continuously improved. A short list of 
ongoing activities includes:
\begin{enumerate}
\item Production of heavy (strange, charm) flavor:
\begin{enumerate}
\item in (thermal) QGP;
\item in parton and string models;
\item confined/deconfined  state comparisons.
\end{enumerate}
\item Evolution of hadronic matter and formation of hadrons in final state:
\begin{enumerate}
\item collective flow of dense matter,
\item hadronization, 
\item hadrochemistry, 
\item approach to chemical equilibrium.
\end{enumerate}
\item Analysis of experimental results:
\begin{enumerate}
\item $m_\bot,\,y$--spectra;
\item particle abundances;
\item HBT correlation analysis.
\end{enumerate}
\end{enumerate}
As indicated above in (i), there are several 
distinct classes of theoretical approaches to
strangeness production in high energy nuclear collisions. A simple
classification arises dividing the models into progressively more
microscopic approaches:
\begin{itemize}
\item[] {\it local thermal and chemical equilibrium models},
\item[] {\it local thermal equilibrium and kinetic description
 of chemical reactions},
\item[] {\it fully kinetic (microscopic) reaction models}.
\end{itemize}
 Within  the local kinetic (temperature)
equilibrium description, allowance must be
made for associated  spatial and temporal
distribution of temperature (energy density), 
and  chemical potentials, and also
the collective flow velocity.
Among the local equilibrium descriptions, 
I distinguish between those based on 
\begin{itemize}
\item[] {\it hadronic gas constituents}\hspace*{1cm} and those based on 
\item[] {\it quark-gluon matter}. 
\end{itemize}
In both cases there are different levels of treatment of chemical 
equilibration:
\begin{enumerate}
\item[] {\it assume relative equilibrium, but not `absolute'  
chemical equilibrium},
\item[] {\it treat only heavy (strange) quarks using kinetic theory},
\item[] {\it  use kinetic theory for all chemical processes}.
\end{enumerate}
When full chemical kinetic theory is employed, we indeed 
have gone half way to microscopic models of collision dynamics, which do 
not assume thermalization of constituents. But even in fully microscopic 
descriptions there are several distinct classes,  
we distinguish two primary approaches: 
\begin{itemize}
\item[] {\it hadronic cascades}
involving {\it reactions between individual hadrons} and 
\item[] {\it quark cascades}, 
which may be further differentiated as  
\begin{itemize}
\item[] {\it color string models} and 
\item[] {\it parton (cascade) models.} 
\end{itemize}
\end{itemize}
This short classification of the theoretical approaches
is in no way complete, as practically all types of 
hybrid models have also been 
proposed. For example,  models which combine 
a smooth transition from confined to deconfined structure, or which 
address kinetic theory  of heavy quarks in the thermal background of
light quarks and gluons. 

Why do we need so many different models? 
Nobody doubts that microscopic models are superior to statistical 
models (and hence is this what we all should be doing?), 
but some of us  also realize that, in practice, microscopic approaches 
suffer critically from the need to understand and be able to 
model all relevant and  accessible reaction mechanisms, including the 
novel phenomena that are yet to be discovered. Thus practice calls for
compromise solutions, in which some aspects are considered 
to be sufficiently precisely described in a simpler (statistical)
manner. Moreover,
the degree to which the reaction is treated as a quantum mechanical process,
as compared to classical two particle cascades, is not understood in principle.
In fact, even if we had much greater computing power available allowing us
to compute within fundamental theoretical approaches such as 
lattice-QCD \cite{Lae96},
we would not know how to treat the dynamics of the collision, the transition
from quantum to classical dynamics, the non-equilibrium aspects and many 
other issues. So the multitude of approaches really reflects on the
exploratory character of the research we are engaged in. 

Each model has its pros and cons: 
microscopic models cannot be used to interpret results without
fine tuning the various implicit and explicit reactions. The
thermal models suffer not only from the perception that kinetic equilibrium is 
introduced {\it deus ex machina}, without proper understanding of the 
dynamics and time scales involved. However, 
the reaction dynamics in local equilibrium 
models are usually relatively simple and often allow one to come to
model-independent conclusions about suitably defined observables. 
However, before using a statistical description we must scrupulously test
which quantities can be treated by near-equilibrium methods. This
often requires the tools of a microscopic model. So both approaches
are indeed complementary.

\section{Quo Imus: Where are we going?}
Quark-gluon plasma is, by the meaning  of these words, a thermally equilibrated  
state  consisting of mobile, color charged quarks and gluons. There is
no requirement for chemical equilibration, in which quark and gluon
abundances are fixed to the Stefan-Boltzmann limit. 
In laboratory experiments only a very short-lived 
QGP phase can be established, and thus by the nature of the circumstance
we should expect, if at all, a deconfined state in a chemical non-equilibrium. 
Moreover, entropy and total strangeness excesses 
are the global observables of deconfinement. There is the 
fundamental issue if thermal equilibrium can be attained in the short
time available. This is a very controversial question, which we would
like to address by inspection of experimental results, which seem to be
strongly in favor. 

The chemical  analysis of  (mostly strange) hadrons produced 
in high energy nuclear reactions offers opportunity for:
\begin{enumerate}
\item a precise determination of the overall strangeness yield;
\item determination,  if for some rarely produced particles the chemical equilibrium is 
approached from below or above, the latter case pointing to deconfinement;
\item an assessment, if strangeness excess is accompanied by  entropy excess 
as would be expected for deconfined source at hadronization. 
\end{enumerate}

An important aspect here is that the specific yield of strangeness produced 
seems to fall rapidly with decreasing available collision energy, or increasing 
reaction volume (or both). The lowest yield seems
to occur in the 158 A GeV Pb--Pb reactions, where a detailed study of 
the phase space occupancy suggests specific abundance at the level of 
$\bar s/b=0.6$--0.7\,.  Same type of analysis yields nearly 50\% higher abundance 
at the level of $\bar s/b=0.9$--1.0 for S--W/Au/Pb 200 A GeV collisions. Yet 
higher yields are suggested by S--S 200 A GeV results, though details are
obscured by the strong longitudinal flow present in this light, yet smaller system
 --- the available  energy is here  still higher since the collision is
symmetric.
\begin{itemize}
\item Could it be that this rapid rise in strangeness yield with collision 
energy occurs since the current SPS experiments occur just at the threshold 
to deconfinement? Alternatively:
\item Is it possible that this rapid change in strangeness yield is 
an intricate effect of the dynamics of highly compressed dense quark matter
arising in particular from  differences in the lifespan of the high 
temperature deconfined phase?
\end{itemize}
We need to clarify this important issue. 

Detailed interpretation of experimental strange flavor flow results 
requires almost always an understanding of the general reaction dynamics
expressed by baryon and meson spectra. These measurements in fact give
us an idea about the stopping of energy and baryon number in the central 
fireball and the entropy (particle multiplicity) yield per participant.
This study can be combined  with a comprehensive analysis of the 
abundance freeze-out conditions of strange  hadrons. 
In such an analysis chemical non-equilibrium 
features for both strange  and non-strange particle production
need to be included. In that way
one obtains a snap-shot of the dense matter fireball, taken  when strange
hadrons stop changing in number,  i.e., at the chemical 
freeze-out. 

Beyond the chemical properties, the local surface 
rest-frame temperature $T$ and local collective
flow velocity $\vec v_{\rm c}$ characterize the  momentum space distribution 
of particles emerging and can be studied in such final state analysis.
However, particles of similar mass and cross section with the 
background (compatible  particles) experience
similar drag  forces arising from the local flow of matter and hence
the ratio of their abundances in some limited region of phase space,
for not too small momenta,
is expected to remain similarly altered by $\vec v_{\rm c}$. While the
surface vector flow $\vec v_{\rm c}$ is a priori largely unknown, 
one simple hypothesis is radial explosion with the transverse to 
collision axis surface velocity reaching the `sound' velocity 
of massless matter, $c/\sqrt{3}$. This discussion shows 
that the theoretical analysis, even in the statistical approach 
is as tedious as is the  experimental data analysis. In conclusion, the
theoretical study of the  great wealth of experimental hadronic 
data is not easy, but there has been considerable progress 
made in recent years, with results becoming more reliable and consistent
between the different groups and strategies of approach.

\begin{figure}[tb]
\vspace*{1.7cm}
\centerline{\hspace*{3.7cm}
\psfig{width=10cm,figure=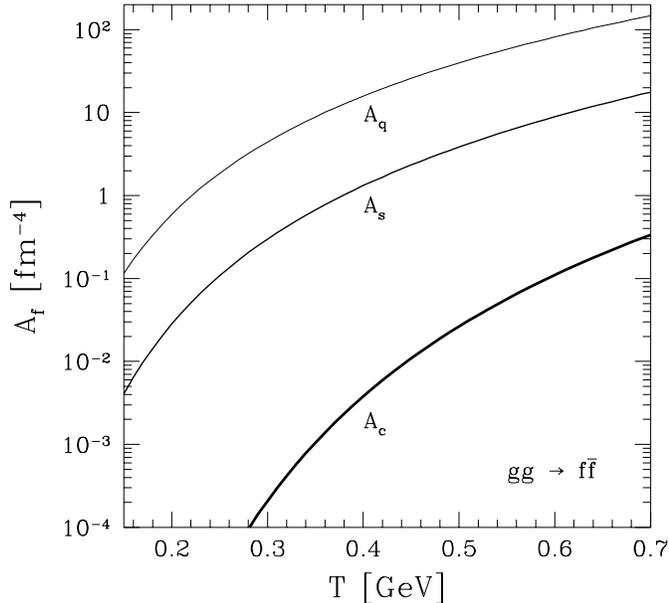}
}
\vspace*{-0.5cm}
\caption{ 
Invariant thermal rate $A_f$ per unit time and volume 
as function of temperature $T$ for production of 
quark pairs in gluon fusion processes $gg\to f\bar f$\,, where 
$f$ is either the light quark $q=u,d$ flavor (thin line, 
$m_q(1\,\mbox{GeV})=10$\,MeV), 
or strange $s$ ($m_s(1\,\mbox{GeV})=200$\,MeV) or
charm $c$ quark flavor (thick line, $m_c(1\,\mbox{GeV})=1.5$\,GeV);
after \cite{LR98}.
\label{Aqsc}
}
\end{figure}

The information we  extract tells us, 
in principle, only about the  momentary  physical properties
of the dense fireball.  However, even such a snap-shot taken 
at the end  of the chemical evolution 
contains clear information about earliest moments of the collision:
the total yield of heavy flavor
is primarily determined in the initial stages of the collision. 
Since it is mainly produced by chemically equilibrated  gluons, 
in particular charm, and to a lesser degree also strangeness 
yield is determined by
the initial temperature.  The situation is best understood inspecting
figure~\ref{Aqsc}, were the differential rate of production of 
flavor, $f=q,s,c$, is shown as a function of temperature, for 
the dominant gluon fusion process. The interaction strength 
of the perturbative QCD  is gauged at the  decay of the 
Z$_0$ particle, and is not a free parameter in this calculation. 
However, for strangeness (but not for charm) the uncertainty in 
$m_{\rm s}(1\,\mbox{GeV})$ translates into an uncertainty of 25\% in
the production rate shown. 
We see in figure~\ref{Aqsc} \cite{LR98}\,, that the expected 
rate of charm production at $T=500$\,MeV
is nearly equal to the rate of strangeness production at  $T=200$\,MeV:
specifically 
$$A_c(500\,\mbox{MeV})\simeq 0.03/\mbox{fm}^4\simeq A_s(200\,\mbox{MeV})\,.$$
This implies that the total yield of charm at RHIC will be as abundant 
as that of strangeness at AGS. However, per hadronic particle produced it 
will be 16 times smaller, considering that the entropy content scales
with third power of $T$\,. 

We conclude that open charm is  more than strangeness 
a `deeply penetrating' probe
of the early QGP conditions, much akin to 
dilepton and direct photon signals. These 
electromagnetic signals are, unlike heavy flavor, 
`deeply hidden' in the general reaction background, and
indeed even the decay products of open charm provide here
considerable background.  However, this can be a blessing: 
the reported excess of the dilepton yield below the $J/\!\Psi$ resonance
has been shown at this meeting to be consistent with the 
decay of open charm mesons \cite{Sco98}, a very tantalizing suggestion 
that is likely to stimulate the effort for direct measurements
of charmed meson decays, in principle possible in the NA57 experiment
at SPS (successor to WA97), and we are looking forward to copious 
appearance of this  flavor observable at RHIC energies, and its 
use as a deeply penetrating diagnostic tool.

We thus have, in the absolute yields 
of strangeness and charm, a measure of the initial stage of the 
collision and in the relative 
hadron yields a snap-shot picture of the freeze-out 
conditions.  Can this information  illuminate the issue of 
deconfinement? Definitively so,
if we are able also to obtain many different snap-shots of the chemical 
freeze-out,  varying energy of colliding nuclei, 
and the participating amount of matter.

Let me close with a few general remarks: 
I am not able to look further than the next 
two years into the future, given the 
phase transition the community is undergoing presently: the exploratory fixed target 
programs at SPS and AGS are coming to an end while the dedicated measurements of the 
properties of deconfined phase are likely to begin at RHIC, where in fact the 
physics is completely unknown. However, 
the discussion we have presented provides for a clear shopping list, which
we will need to fill in next two years.  My personal experimental and theoretical 
priorities are as follows: 

\section*{\bf Experiment}
\begin{enumerate}
\item  Complete analysis of SPS-158 A GeV, with the aim of presenting
both particle spectra and precise relative hadronic/strange particle yields; 
\item Determine the yield of strange antibaryons at the AGS
Au-beam.
\item Obtain strange particle yields, and spectra, at the energy intermediate between
AGS and maximum available at SPS, and compare with the lower/higher energy
results in comparable conditions.
\item Ramp the energy of the SPS up to check if strangeness yield 
rises even with a (small) energy increase. Vary the size of projectile/target
combination (e.g. using Ag--Ag collisions or/and trigger condition on heavy
projectile/target combination) to see how energy and reaction
volume combine to determine the overall strangeness yield.
\item Obtain first results on reaction mechanisms and 
strangeness production (Kaons) at the 
12 times higher energies becoming  available at RHIC. 
\item In view of the likely phase-out of experimental research programs at
 AGS and SPS,
and the likely occurrence of the deconfinement transition below the 
lower RHIC energy limit (4 times the current SPS-CM-available energy) 
commence development  of a  nuclear intersecting beam collider  ranging 
the deconfinement transition region. 
\end{enumerate}
\section*{\bf Theory}
\begin{enumerate}
\item Complete development of consistent analysis 
programs of the experimental data
based on  competing reaction mechanisms (confinement/deconfinement) 
and theoretical approaches (thermal/kinetic/parton).
\item Refine the understanding of the hadronization mechanisms of 
quark-gluon plasma and the production of strange hadrons at phase transition
conditions.
\item Develop the equations of state of hot and dense thermal matter in confined and
deconfined conditions without assumptions about chemical equilibrium.
\item Continue progressing towards an understanding of QCD-vacuum structure and 
properties of the phase transition at finite baryon density.
\end{enumerate}
I am looking forward to our next meeting to be held in July 2000. 
All welcome to {\bf Strangeness 2000--USA!}

\section*{Acknowledgments}
I would like to thank U. Heinz, J. Letessier, and E. Quercigh 
for the careful reading and valuable comments about the contents of this
manuscript, and  collaborations NA35/NA49 and WA97 for the permission to use 
their figures.  This work was supported
by a grant from the U.S. Department of
Energy,  DE-FG03-95ER40937\,.



\section*{References}

\begin{thebibliography}{99}\small
\setlength{\itemsep}{-.015cm}

\bibitem{S98} Proceedings of the International Conference: 
{\it Strangeness in Quark Matter 1998}, held in Padova, 
July 20--24, 1998, to appear in
{\it J. Phys.} G (1999), M. Morando et al., Eds.

\bibitem{Raf80} 
J. Rafelski,  pp 282--324, 
GSI Report 81-6, Darmstadt, May 1981;
Proceedings of the Workshop on {\it Future Relativistic 
Heavy Ion  Experiments}, held at GSI, Darmstadt, 
Germany, October 7--10, 1980, R. Bock and R. Stock, Eds.,
(see in particular section 6, pp 316--320); see also:\\
J. Rafelski,  pp 619--632 in {\it New Flavor and Hadron Spectrosopy}, 
Ed. J. Tran Thanh Van (Editions Frontiers 1981), 
Proceedings of XVIth Rencontre de Moriond -- Second Session, 
Les Arcs, March 21--27, 1981;\\
J. Rafelski, {\it Nucl. Physics} A {\bf 374}, 489c (1982) --- Proceedings
of ICHEPNC held 6--10 July 1981 in Versailles, France.

\bibitem{RH81} J. Rafelski and R. Hagedorn, in: {\it Statistical
Mechanics of Quarks and Hadrons}, H. Satz, Ed., North Holland, 
(Amsterdam 1981) p.\,253.

\bibitem{BZ82}
 T.S. Biro and  J. Zimanyi,
{\it Phys. Lett.} B {\bf 113}, 6 (1982);
{\it Nucl. Phys.} A {\bf 395},525 (1983).

\bibitem{RM82}
{J. Rafelski and B. M\"uller}, {\it Phys. Rev. Lett}
{\bf 48}, 1066 (1982); {\bf 56}, 2334E (1986);
J. Rafelski, {\it Phys. Rep.} {\bf 88}, 331 (1982).

\bibitem{Raf84}
J. Rafelski, {\it Nucl.  Phys.} A {\bf 418}, 215 (1984).

\bibitem{KMR86}
{P.~Koch, B.~M\"uller and J.~Rafelski},
{\it Phys. Rep.} {\bf 142}, 167 (1986).

\bibitem{BCDH95} 
N. Bili\'c, J. Cleymans, I. Dadi\'c and D. Hislop,
{\it Phys. Rev.} C {\bf 52}, 401 (1995).
 
\bibitem{LTR96}
J. Letessier, J. Rafelski,  and A. Tounsi, 
{\it  Phys. Lett. }B {\bf 389},  586 (1996). 

\bibitem{NA49Jer}
D. R\"ohrig,  for the NA49 Collaboration, 
``Recent results from NA49 experiment on 
Pb--Pb collisions at 158 GeV per nucleon'',
Jerusalem 1997, to appear in proceedings;
NA49 note 145, available at:
http://na49info.cern.ch/cgi-bin/wwwd-util/NA49/NOTE?145.

\bibitem{Gab98}
Frank Gabler,  for the NA49 Collaboration, in this volume.

\bibitem{Eva98}
David Evans, for the WA85 and the WA94 Collaborations,
in this volume.

\bibitem{Cal98}
R. Caliandro, for the WA97 collaboration,
in this volume, and private communications.

\bibitem{Czi98}
P. Csizmadia, P.Levai, S.E. Vance, T.S. Biro, W. Guylassy, 
and J. Zim\'anyi, in this volume.

\bibitem{ISR84}
T. \AA kesson et al., AFS-collaboration, 
{\it Nucl. Phys.} B {\bf 246}, 1 (1984).

\bibitem{SppS89}
R.E. Ansorge  et al., UA5-collaboration, 
{\it Nucl. Phys.} B {\bf 328}, 36 (1989).

\bibitem{ISRalfa}
T. \AA kesson et al., AFS-collaboration, 
{\it Phys. Rev. Lett.} {\bf 55}, 2535 (1985).

\bibitem{Kin98}
J.B. Kinson, in this volume.

\bibitem{LRT96b} 
{J. Rafelski, J. Letessier and A. Tounsi},
{\it Acta Phys. Pol.} B {\bf 27}, 1035 (1996), and references therein.

\bibitem{Mar98}
Spiros Margetis, for the WA49  Collaboration, in this volume,
and personal communication.

\bibitem{WA97LietPad}
R. Lietawa for the WA97 Collaboration, in this volume.

\bibitem{Gan98}
R. Ganz, for the E917  Collaboration, in this volume.

\bibitem{Chu98}
P. Chung, for the E895  Collaboration, in this volume.

\bibitem{Lae96} 
E. Laermann, {\it Nucl. Phys.} A {\bf 610},{1c} (1996).

\bibitem{LR98}
J. Letessier and J. Rafelski, ``Charmed Particle Signatures of 
Deconfinement'', in preparation.

\bibitem{Sco98}
E. Scomparin, for the NA50 Collaboration, in this volume.

\end{thebibliography}
\end{document}